\documentclass[twocolumn,english,prb]{revtex4-1}

\usepackage[T1]{fontenc}
\usepackage[latin9]{inputenc}
\usepackage{amsmath}
\usepackage{graphicx}

\makeatletter

\providecommand{\tabularnewline}{\\}

\makeatother

\usepackage{babel}
\begin{document}
\title{Dopant levels in large nanocrystals using stochastic optimally tuned
range-separated hybrid density functional theory}
\author{Alex J. Lee}
\email{ajemyunglee@lbl.gov}

\affiliation{Department of Chemistry, University of California and Materials Science
Division, Lawrence Berkeley National Laboratory, Berkeley, CA 94720,
USA}
\author{Ming Chen}
\affiliation{Department of Chemistry, University of California and Materials Science
Division, Lawrence Berkeley National Laboratory, Berkeley, CA 94720,
USA}
\author{Wenfei Li}
\affiliation{Department of Chemistry and Biochemistry, University of California,
Los Angeles, CA 90095, USA}
\author{Daniel Neuhauser}
\affiliation{Department of Chemistry and Biochemistry, University of California,
Los Angeles, CA 90095, USA}
\author{Roi Baer}
\affiliation{Fritz Haber Research Center for Molecular Dynamics and Institute of
Chemistry, The Hebrew University of Jerusalem, Jerusalem 9190401,
Israel}
\author{Eran Rabani}
\affiliation{Department of Chemistry, University of California and Materials Science
Division, Lawrence Berkeley National Laboratory, Berkeley, CA 94720,
USA}
\affiliation{The Raymond and Beverly Sackler Center for Computational Molecular
and Materials Science, Tel Aviv University, Tel Aviv 69978, Israel}
\begin{abstract}
We apply a stochastic version of an optimally tuned range-separated
hybrid (OT-RSH) functional to provide insight on the electronic properties
of P- and B- doped Si nanocrystals of experimentally relevant sizes.
We show that we can use the range-separation parameter for undoped
systems to calculate accurate results for dopant activation energies.
We apply this strategy for tuning functionals to study doped nanocrystals
up to $2.5$~nm in diameter at the hybrid functional level. In this
confinement regime, the P- and B- dopants have large activation energies
and have strongly localized states that lie deep within the energy
gaps. Structural relaxation plays a greater role for B-substituted
dopants and contributes to the increase in activation energy when
the B dopant is near the nanocrystal surface.
\end{abstract}
\maketitle

\section{\label{sec:Introduction}Introduction}

Understanding the properties of dopants in semiconductor nanostructures
is a crucial issue for technological applications since it is often
the dopants that functionalize a device and control its desired properties.\citep{Ossicini2006,Norris2008,Mocatta2011,Amit2013,Hori2016,Li2018,lau2019quantum}
In particular, doped silicon quantum dots have shown promise in photovoltaic
and photonic applications due to their size tunability and processability.\citep{Mavros2011,Hamid2013,Somogyi2017}
A key dopant property is its activation energy, which often behaves
differently in the nanoscale. For example, when P- (phosphorus) and
B- (boron) dopants are introduced to bulk Si, they create shallow
impurity states that can act as donors/acceptors of charge carriers.
However, in nanocrystals such dopant impurities become deep states
due to quantum confinement and dielectric mismatch. 

Numerous tools are available to describe dopant properties for extended
systems with periodic boundary conditions in a supercell. The state-of-the-art
approach is based on a combination of density functional theory (DFT)
with many-body perturbation theory (MBPT), typically within the so-called
``GW'' approximation.\citep{Hybertsen1985,Hybertsen1986} However,
the application of the GW method to large confined systems such as
nanocrystals is limited by the steep scaling and the slow convergence
with respect to empty states.\citep{Rohlfing2000} Furthermore, the
description of optical excitations requires the use of the Bethe-Salpeter
equation (BSE), which scales even steeper with system size, limiting
its application to small systems.

In recent years, we have developed a set of stochastic orbital techniques~\citep{Baer2013,Neuhauser2013,Neuhauser2013-2,Ge2014,Neuhauser2014,Gao2015,Neuhauser2015,Rabani2015,Vlcek2016,Vlcek2018,Vlcek2018-2,Chen2019,Li2019}
which significantly reduce the scaling and computational costs of
both the GW and BSE approaches by introducing a controlled statistical
error in the calculated observables. This enables the application
of both methods to extremely large, experimentally relevant system
sizes containing thousands of electrons.\citep{Neuhauser2014,Rabani2015,Vlcek2016}
Some applications, like those involving linear response time-dependent
DFT for optical excitations,\citep{Rabani2015,Vlcek2019,Zhang2020}
require the use of a quasiparticle model Hamiltonian, which is not
available through the GW method.

Density functional theory has been a major tool for quasiparticle
electronic structure calculations, but its local and semi-local approximations
poorly predict electronic properties like the fundamental gaps, ionization
energies, and electron affinities. This is especially a problem for
dopant properties since local approximations can erroneously predict
shallow dopant levels when they are in fact deep.\citep{Rurali2009,Niquet2010,Lee2014}
Optimally-tuned range-separated hybrid (OT-RSH) functionals offer
a solution to this problem.\citep{Livshits2007,Baer2010} Specifically,
the optimally tuned Baer-Neuhauser-Livshits (BNL) functional\citep{Baer2005,Livshits2007}
has been shown to provide an accurate description of the fundamental
band gaps for molecules, nanocrystals, and bulk materials.\citep{Stein2010,Refaely2011,Kronik2012}
One of the major strengths of this approach\citep{Salzner2009} is
that it can be tuned to the system of interest based on a physical
constraint, avoiding the use of empirical fitting parameters that
might not reproduce the correct physics. In fact, it has been suggested
that hybrid functionals tuned to physical constraints give some of
the most reliable electron densities within DFT.\citep{Medvedev2017}

One of the questions in applying OT-RSH functionals to study dopant
properties is this: How do we tune the functional to correctly calculate
dopant energies and at the same time maintain accuracy in describing
the band structure? Oftentimes, we want to study many different dopant
types and locations. Here, the system-specific tunability that is
one of OT-RSH's strengths becomes a liability. Having to repeat the
tuning procedure for every dopant structure would quickly become cumbersome
and resource intensive. 

In this paper, we apply a stochastic formulation
of the optimally tuned BNL range-separated hybrid functional~\citep{Neuhauser2015}
to study dopant properties in silicon nanocrystals of up to $1600$
electrons. We show that the range-separation parameter for undoped
systems also gives accurate results for dopant activation energies.
We thus provide a strategy for tuning functionals for doped systems.
This strategy may be generalizable for many different dopant types
and positions within the nanocrystal. We demonstrate the usefulness
of this strategy in conjunction with stochastic techniques to provide
insight on dopant properties for nanocrystals of experimentally relevant
sizes.

\section{Theory and computational details}

We summarize the main points in the theory of optimally-tuned range-separated
hybrid functionals and its stochastic formulation. Consider a zero-temperature
ensemble for a system with an average number of electrons equal to
$N-x$ , where $N$ is an integer and $x\in\left(0,1\right).$ For
this system, the energy curve $E\left(N-x\right)$ should be linear
in $x$, so the slope is equal to the negative of the ionization energy,
i.e. to $E\left(N\right)-E\left(N-1\right).$ A similar condition
holds for $E\left(N+x\right)$ which should be linear in $x$ with
a slope equal to the negative of the electron affinity, $E\left(N+1\right)-E\left(N\right)$. 

In exact Kohn-Sham (KS) DFT the ionization energy corresponds to the
negative of the HOMO energy. Thus, the line $E\left(N-x\right)$ should
have a slope equal to the HOMO energy of the $N$ electron system,
and the line $E\left(N+x\right)$ should have a slope of the HOMO
for the $N+1$ system. That is, in KS-DFT the LUMO of the $N$ electron
system is \emph{not} equal to the HOMO of the $N+1$ system. The difference
is due to the \emph{derivative discontinuity }in the exchange-correlation
energy functional as the number of electrons goes from slightly below
$N$ to slightly above it.\citep{Perdew1982,Perdew1997,MoriSanchez2008,Yang2012,MoriSanchez2014}
This behavior of the exact KS functional is not reproduced correctly
by local or semi-local KS functionals, such as LDA and the various
types of GGAs where the functional exhibits no derivative discontinuity.
To compensate for the lack of derivative discontinuity, the energy
$E(N\pm x)$ becomes non-linear.\citep{stein2012curvature} 

One way to account for this lack and for the non-linearity in $E(N\pm x)$
is to use optimally-tuned range-separated hybrid functionals within
generalized Kohn-Sham DFT (GKS-DFT).\citep{Livshits2007,Baer2010,Kronik2012}
Specifically, we employ the range-separated hybrid functional following
the proposal by Savin to use full-exchange at long distances.\citep{savin1995beyondthe,iikura2001alongrange,Toulouse2004,Baer2005}
The exchange term is divided into long- and short-range components
\begin{equation}
\frac{1}{r}=\frac{\text{erf}\left(\gamma r\right)}{r}+\frac{\text{erfc}\left(\gamma r\right)}{r}
\end{equation}
where $\gamma$ is the range-separation parameter that controls the
distance upon which the potential is switched from long to short range,
$\text{erf}\left(x\right)$ is the error function, and $\text{erfc}\left(x\right)$
is the complimentary error function. The long-range term is calculated
explicitly through a Fock-like exchange operator while the short-range
term is approximated by a screened local exchange functional. 

This hybrid construction is attractive because it maintains the correct
long-range asymptotic behavior, decaying as $1/r$, whereas local
exchange functionals are known to decay too rapidly. This property
allows a full fraction of exact exchange to cancel out the long-range
self-interaction error in the Hartree energy part of the DFT functional.
A complication is that the range-separation parameter $\gamma$ is
in principle a functional of the density that we currently do not
know how to construct. This is where the approach of \emph{optimally
tuning} becomes useful.\citep{Livshits2007,Baer2010} 

The tuning of
$\gamma$ imposes a physical constraint to the system instead of relying
on universal or empirical fittings. The physical constraint ensures
that it satisfies the linearity of the ensemble energy $E\left(N\pm x\right)$
with a fractional number of electrons $x$. Since by Janak's theorem\citep{Janak1978}
the energy slope is equal to the orbital energy, the constant slope
requirement is equivalent to 
\begin{equation}
\frac{\partial\varepsilon_{H/L}}{\partial f_{H/L}}=0
\end{equation}
where $\varepsilon_{H/L}$ refers to the HOMO or LUMO orbital energy
and $f_{H/L}$ is its occupancy. Hence, the approach explicitly constructs
a GKS functional such that the IP corresponds to the HOMO and the
EA to the LUMO as closely as possible, meaning the functional should
be able to accurately describe fundamental band gaps. In many cases,
this approach works well  and can predict fundamental gaps close to
experiment and to MBPT methods for various atomic and molecular systems,\citep{Stein2010,Refaely2011}
as well as Rydberg\citep{Baer2010} and charge-transfer excitations
(within TDDFT).\citep{stein2009reliable,Kronik2012}

Dopant activation energies are calculated similar to fundamental gaps.
For electron donors, the activation energy is defined as the energy
difference to ionize the dopant and place the electron back into the
undoped structure. For acceptors, it is the difference to remove an
electron from the undoped structure and place it back into the acceptor
level. To summarize,

\begin{align}
E_{\text{act}}=IP_{d}-EA_{u} & \,\,\,\,\,\,\,\text{[donor]}\nonumber \\
E_{\text{act}}=IP_{u}-EA_{d} & \,\,\,\,\,\,\,\text{[acceptor]},
\end{align}
where the subscripts $d$ and $u$ refer to the doped or undoped structures.
We can ensure the dopant IPs and EAs are calculated correctly by tuning
the functionals to the dopant levels explicitly. However, repeating
the tuning procedure becomes costly, especially since one often wants
to examine many different dopant types at different dopant locations.
One of the key findings of the paper is that we can bypass this tuning
step and simply use $\gamma$ for the undoped structure in the spirit
of using one constant $\gamma$ for the fundamental gap. In this case,
the dopant can be viewed as a perturbation to the electronic structure
of the undoped system and does not significantly affect its electronic
environment.

A roadblock to using hybrid functionals is that they are much more
expensive to use than local multiplicative functionals like the local
density approximation (LDA) since they involve the explicit calculation
of the orbital-dependent exchange energy. In a ``deterministic''
(non-stochastic) calculation, the long-range exchange operator is
given by the expression:

\begin{equation}
\hat{K}_{x}^{lr}[\psi_{i}(\boldsymbol{r})]=-\sum_{j=1}^{N_{\text{occ}}}\psi_{j}(\boldsymbol{r})\int d\boldsymbol{r}'\psi_{j}^{*}(\boldsymbol{r}')\psi_{i}(\boldsymbol{r}')V_{\text{c}}^{\gamma}(|\boldsymbol{r}-\boldsymbol{r}'|)\label{eq:4}
\end{equation}
where $V_{\text{c}}^{\gamma}(|\boldsymbol{r}|)$ is the long-range
screened Coulomb potential governed by the range parameter $\gamma$.
The computation of this exchange term roughly scales quadratically
as $N_{\text{occ}}N_{\text{grid}}$, where $N_{\text{occ}}$ is the
number of occupied states and $N_{\text{grid}}$ is the size of the
basis set (the number of real-space grid points), both of which increase
with system size. In a typical hybrid functional application, this
exchange term takes up the overwhelming majority of the total computation
time.

Stochastic techniques have been developed to overcome the deterministic
scaling limit to make hybrid functional calculations tractable for
large systems.\citep{Neuhauser2015} Using a technique called stochastic
resolution of the identity, we can decouple the $\mathbf{r}$ and
$\mathbf{r'}$ indices in Eq.~(\ref{eq:4}). We introduce a stochastic
orbital $\xi(\boldsymbol{r})$ that assigns a random sign to each
real-space grid point ($dV$ is the volume per grid point):

\begin{equation}
\xi(\boldsymbol{r})=\langle\boldsymbol{r}|\xi\rangle=\pm\frac{1}{\sqrt{dV}}.
\end{equation}
We define

\begin{equation}
\eta(\boldsymbol{r})=\sum_{j=1}^{N_{\text{occ}}}\psi_{i}(\boldsymbol{r})\langle\psi_{i}|\xi\rangle,\label{eq:}
\end{equation}
which is the stochastic orbital projected on to the occupied space
spanned by $\psi_{i\in\text{occ}}(\boldsymbol{r})$. Similarly, we
can write the Coulomb potential in a stochastic representation as
\[
\zeta(\boldsymbol{r})=\frac{1}{(2\pi)^{3}}\int d\boldsymbol{G}\sqrt{\tilde{V}_{\text{c}}^{\gamma}(\boldsymbol{G})}e^{i\varphi(\boldsymbol{G})}e^{\text{i}\boldsymbol{G}\cdot\boldsymbol{r}},
\]
where $\tilde{V}_{\text{c}}^{\gamma}(\boldsymbol{G})$ is the Fourier
transform of $V_{\text{c}}^{\gamma}(|\boldsymbol{r}|)$ and $\varphi(\boldsymbol{G})$
is a random phase between $[0,2\pi]$. Now we can rewrite the exchange
term in Eq.~\ref{eq:4} as an average ($\left\langle \cdots\right\rangle _{\xi,\varphi}$)
over the stochastic orbitals:

\[
\hat{K}_{x}^{lr}[\psi_{i}(\boldsymbol{r})]=-\left\langle \eta(\boldsymbol{r})\zeta(\boldsymbol{r})\int d\boldsymbol{r}'\zeta^{*}(\boldsymbol{r}')\eta^{*}(\boldsymbol{r}')\psi_{i}(\boldsymbol{r}')\right\rangle _{\xi,\varphi}.
\]
Defining the product of stochastic orbitals $\chi(\boldsymbol{r})=\zeta(\boldsymbol{r})\eta(\boldsymbol{r})$,
we simplify the above expression to 

\begin{equation}
\hat{K}_{x}^{lr}[\psi_{i}(\boldsymbol{r})]=-\frac{1}{N_{\text{sto}}}\sum_{\chi=1}^{N_{\text{sto}}}\chi(\boldsymbol{r})\langle\chi|\psi_{i}\rangle.\label{eq:6}
\end{equation}
The scaling for the exchange becomes $N_{\text{sto}}N_{\text{grid}}$,
where $N_{\text{sto}}$ is the number of stochastic orbitals. If $N_{\text{sto}}$
goes to infinity, we recover the deterministic result in Eq.~(\ref{eq:4})
exactly. In this sense, $N_{\text{sto}}$ becomes another convergence
parameter that we control at the cost of introducing statistical error.
Remarkably, we find that $N_{\text{sto}}$ is often independent of
system size or can even \emph{decrease} with system size (see Sec.~Results
and Ref.~\onlinecite{Neuhauser2015} for details), so the scaling
for stochastic exchange becomes \emph{quasilinear}.

We implemented the stochastic BNL functional into a plane wave DFT
code to calculate quasiparticle spectra and fundamental gaps for H-passivated
Si nanocrystals ranging from 1 to $2.5$~nm in diameter, containing
up to $N_{e}\approx1600$ electrons. We tune the range separation
parameter $\gamma$ for each nanocrystal size based on the physical
constraint described above. We used a kinetic energy cutoff of $40$~Ry
for the density, which converges band gaps to within $0.1$~eV. We
treat the divergent $G=0$ term in the exchange energy using the Gygi-Baldereschi
method.\citep{Gygi1986} Structural relaxations were performed with
\textsc{Quantum ESPRESSO} with the LDA functional.\citep{Giannozzi2009,Giannozzi2017}
For more computational details, see the Supplementary Material.

\section{Results}

\begin{figure}[t]
\noindent \centering{}\includegraphics[width=0.8\linewidth]{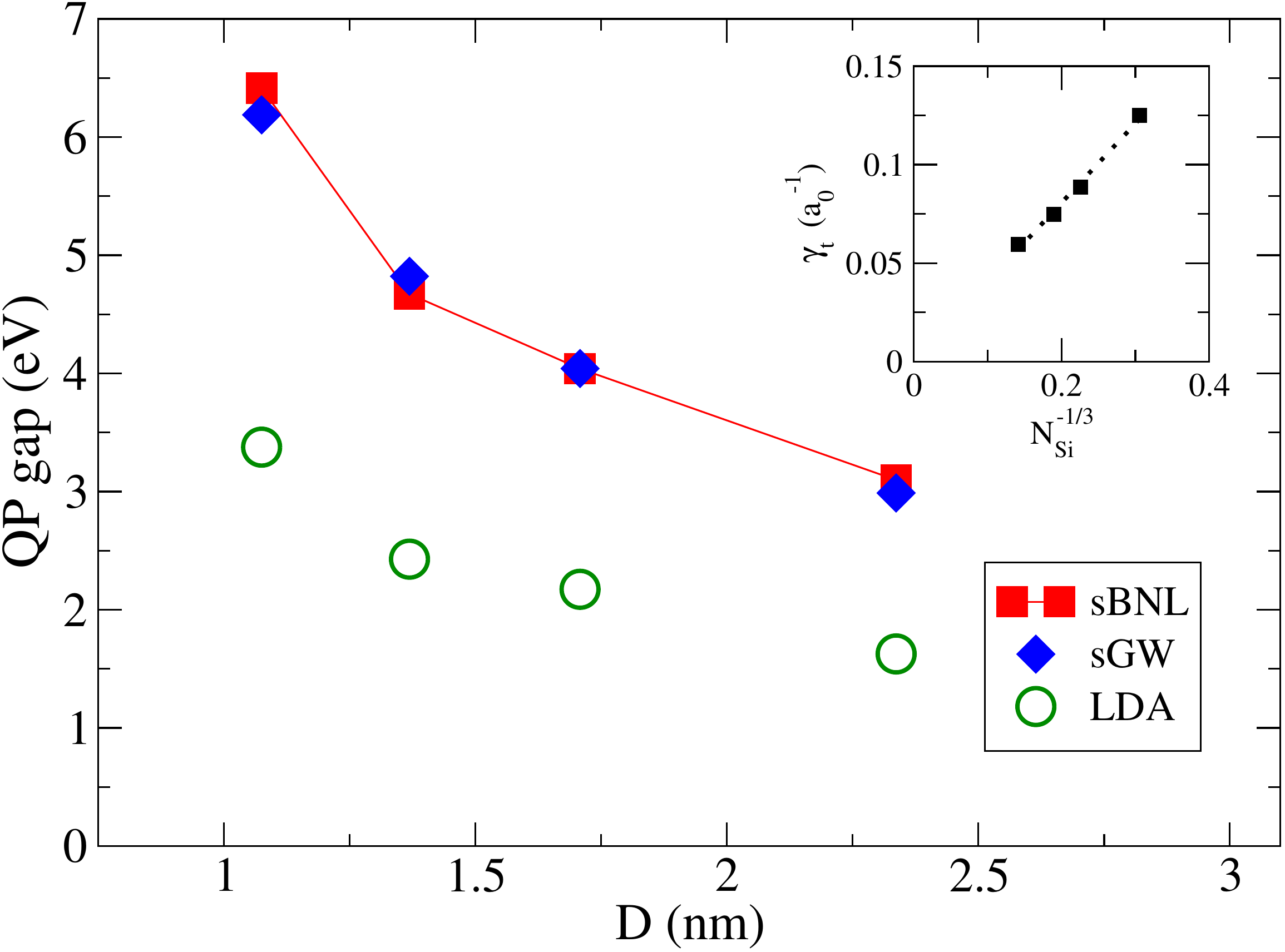}
\caption{\label{fig:1} QP gap comparisons for Si NCs ranging from $1$ to
$2.5$~nm. The sGW results are from Ref.~\onlinecite{Neuhauser2015}.
Inset: The trend in the optimized $\gamma$s for different NC sizes
is nearly linear with respect to $N_{\text{Si}}^{-1/3}$.}
\end{figure}

Using optimized $\gamma$s obtained from the tuning procedure, in
Fig.~\ref{fig:1} we compare quasiparticle (QP) gaps for Si NCs of
different sizes obtained by different theoretical methods. The agreement
between stochastic BNL (sBNL) and stochastic GW (sGW)\citep{Neuhauser2014,Vlcek2016,Vlcek2018,Vlcek2018-2}
is remarkable whereas LDA significantly underestimates the gap. We
note that $\gamma$ decreases nearly linearly with system size (see
inset in Fig.~\ref{fig:1}),\citep{Stein2010,Neuhauser2015} illustrating
the favorable scaling properties of sBNL: Since $\gamma$ becomes
smaller as system size increases, the fraction of stochastic exact
exchange mixed into the functional also becomes smaller. Therefore,
we can use fewer stochastic orbitals to converge the exchange energy
to within an acceptable error, which decreases the computation time. 

Note that if we extrapolate $\gamma$ to the bulk limit ($N_{\text{Si}}^{-1/3}\rightarrow0$),
we would find $\gamma_{\text{bulk}}$ to be nearly $0$, meaning the
exchange is entirely handled by the short-range local functional,
which gives incorrect results for the gap. This problem can be addressed
by introducing a dielectric coefficient to modulate long-range charge
transfer.\citep{Dunietz2018,Kummel2018} Without the dielectric term,
we would expect $\gamma$ to saturate at the value of $\gamma=0.02$
(empirically determined) due to exchange-driven orbital localization.
This phenomenon has been observed with the BNL functional for 1D chains
but has yet to be seen for 3D systems.\citep{Vlcek2016-2} Based on
our model, this localization should occur at around $N_{\text{Si}}=8200$,
or a system with over $N_{e}=32,800$ electrons.

\begin{figure}
\begin{centering}
\includegraphics[width=0.8\linewidth]{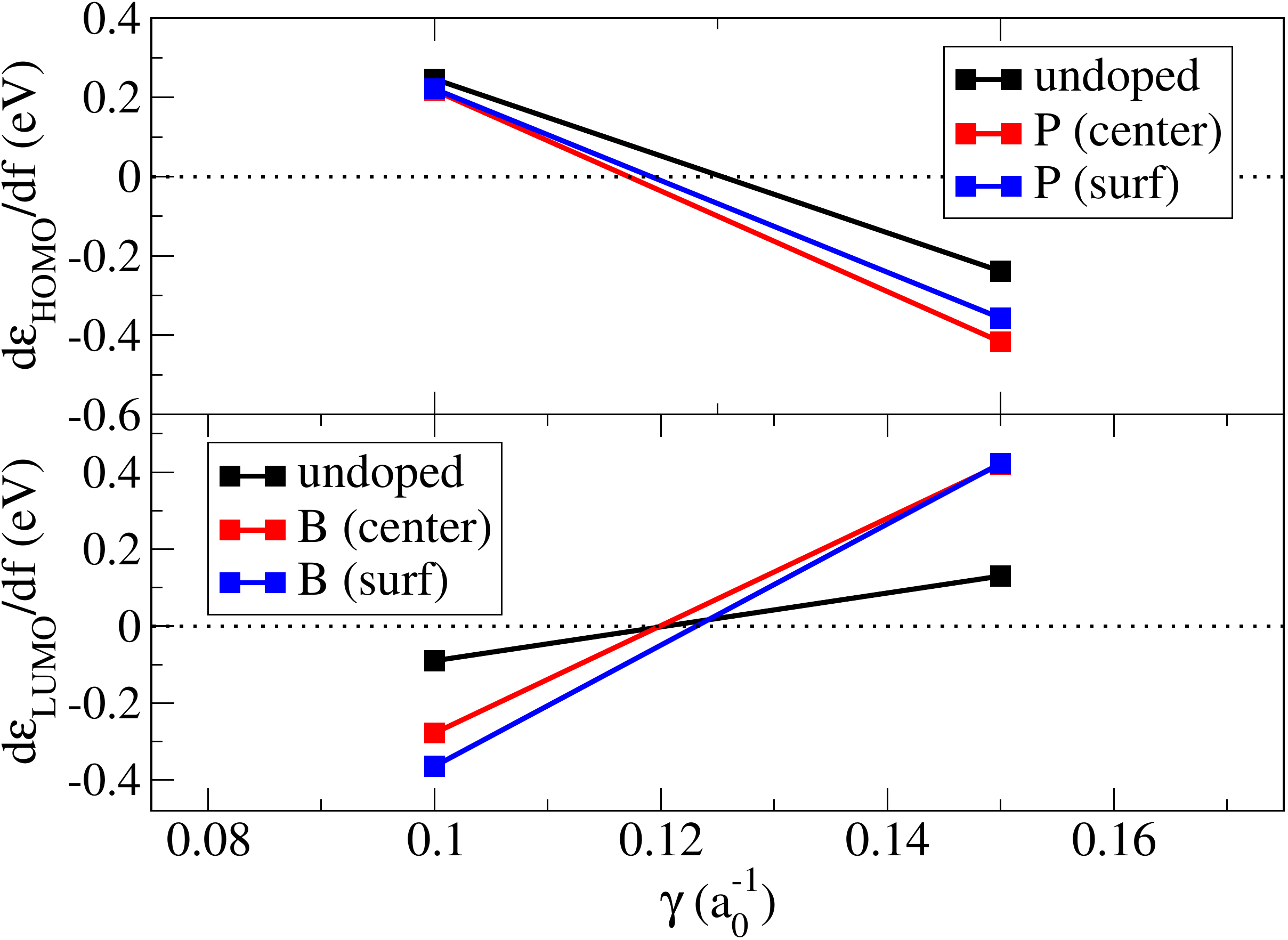}
\par\end{centering}
\caption{\label{fig:2} Tuning $\gamma$ for the undoped and doped structures
in the $D=1$~nm NC. The optimal $\gamma$ values are given by the
intersect with the $y$-axis (dotted line). Upper panel: A partial
charge is removed from the HOMO, which is the dopant level for a P
donor. Lower panel: A partial charge is added to the LUMO, which is
the dopant level for a B acceptor.}
\end{figure}

\begin{table*}
\begin{tabular}{c|cccc}
system & optimized $\gamma$ & energy level & energy level & difference\tabularnewline
 &  & at $\gamma$ & at $\gamma=0.125$ & $\Delta E$\tabularnewline
\hline 
undoped (HOMO) & 0.125 & N/A & N/A & N/A\tabularnewline
undoped (LUMO) & 0.120 & -1.351 & -1.313 & 0.038\tabularnewline
P (center) & 0.117 & -4.355 & -4.387 & -0.032\tabularnewline
P (surface) & 0.119 & -4.336 & -4.360 & -0.024\tabularnewline
B (center) & 0.120 & -4.478 & -4.462 & 0.016\tabularnewline
B (surface) & 0.123 & -4.182 & -4.173 & 0.009\tabularnewline
\end{tabular}

\caption{\label{tbl1:tune} Tuning comparisons for $D=1$~nm nanocrystal.
$\gamma$s are in units of $a_{0}^{-1}$ and energies in eV. The difference
shown is $\Delta E=E(\gamma=0.125)-E(\gamma)$.}
\end{table*}

Next, we show that $\gamma$s tuned for the undoped structures give
reasonable results for doped system properties. We test a P substituted
dopant for an electron donor, and a B substituted dopant for an electron
acceptor in two dopant locations: one at the center of the nanocrystal,
and another near the nanocrystal surface. For the surface dopant,
we substitute a four-coordinated Si atom as far from the center as
possible. We relax the doped structures in all cases. 

In Table~\ref{tbl1:tune} we show tuning comparisons for the smallest
($D=1$~nm) nanocrystal. For this system, we repeat the tuning procedure
to optimize $\gamma$s for each of the doped structures (Fig.~\ref{fig:2}).
This ensures that the physical constraint for the tuning applies directly
to the dopant level itself. When tuning the P-doped structures (donors),
we remove a small partial charge ($+0.05e$) from the HOMO; when tuning
the B-doped structures (acceptors), we add a partial charge ($-0.05e$)
to the LUMO. We plot how much the dopant energies change when using
$\gamma$s tuned specifically for the doped structures compared to
undoped system $\gamma$. We find the change to be negligible, the
largest difference being $32$~meV. This difference is smaller than
the change in the undoped structure energy gap ($38$~meV) when using
one $\gamma$ for the HOMO and LUMO. We conclude that we can also
use one $\gamma$ (the undoped value) to calculate accurate dopant
energies. This strategy likely holds for any four-coordinated dopant
location, from the center of the crystal to the surface.

\begin{figure}[b]
\centering{}\includegraphics[width=0.95\linewidth]{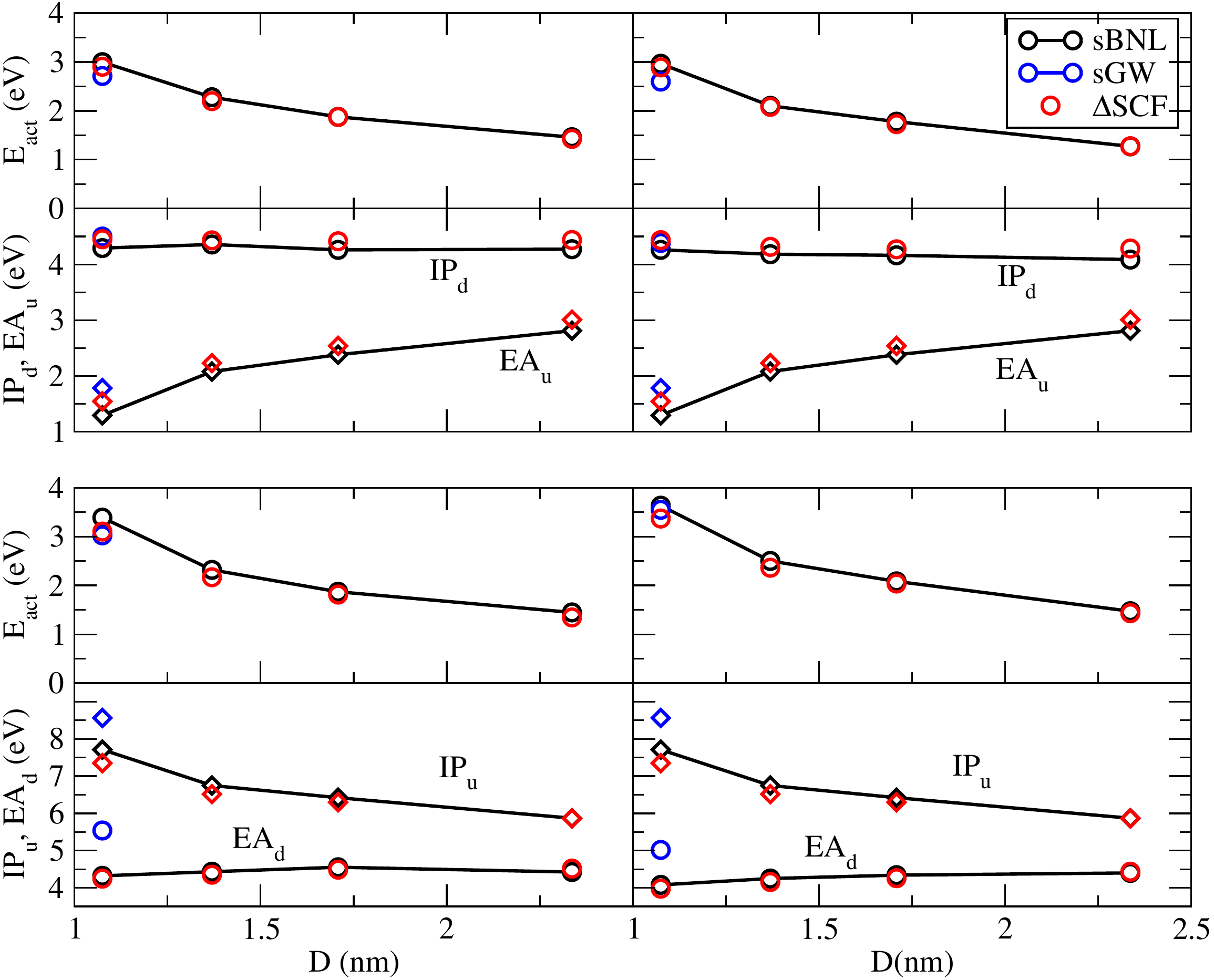}\caption{\label{fig:3} Dopant activation energies for Si NCs with P (upper
panels) and B (lower panels) substituted at the center (left panels)
and surface (right panels). Component IPs and EAs are also shown.
The plotted $\Delta\text{SCF}$ results were performed with the LDA
functional.}
\end{figure}

In Fig.~\ref{fig:3} we plot the dopant activation energies, ionization
energies, and electron affinities for all nanocrystal sizes. For electron
donors like P, we calculate the activation energy as $E_{\text{act}}=IP_{d}-EA_{u}$;
for acceptors like B, we calculate the energy as $E_{\text{act}}=IP_{u}-EA_{d}$,
where the subscripts $d$ and $u$ refer to the doped or undoped structures.
In $\Delta\text{SCF}$, we obtained the IP and EA from total energy
differences of charged ($+1e$ or $-1e$) and neutral calculations.
In the sBNL and sGW calculations, we obtained the IP and EA from the
negative of the corresponding HOMO and LUMO eigenvalue energies (for
sBNL) and associated quasiparticle energies (for sGW, see below).

For the sGW calculations, we obtained good agreement with sBNL
on the smallest cluster ($D=1$~nm) with the $\text{G\ensuremath{\text{W}_{\text{0}}}}$~energies
based on an LDA starting point and with the $\text{G}_{0}$~energies
self-consistently iterated.\citep{Vlcek2018-3} For the larger doped
clusters, the LDA results become essentially metallic, so the $\text{G\ensuremath{\text{W}_{\text{0}}}}\,$energies
are not reliable. Using starting points from other local or semilocal
functionals (PBE or even the SCAN metaGGA functional\citep{Sun2015})
did not improve results. Future studies should examine sGW for dopant
energies based on a better starting point, perhaps from the BNL itself.

As a representative data set, Table~\ref{tbl2:stoGWcompare} compares
dopant activation energies calculated using different theoretical
methods for the $D=1$~nm nanocrystal. As expected, LDA performs
poorly, erroneously predicting the dopant level to be nearly shallow.
We find that sGW tends to predict deeper energy states than sBNL (seen
by the larger IPs and EAs), but the activation energies calculated
from the difference of these states agree well with sBNL.
Similar energy shifts have been observed in other comparative studies,\citep{Stein2010,Refaely2011,Neuhauser2015}
where in some cases the BNL functional agrees with IPs from experiment
better than GW.\citep{Refaely2011} Interestingly, we find that $\Delta$SCF
(over LDA and BNL) gives almost identical results to sBNL. $\Delta$SCF
seems to work particularly well when the added (or removed) charge
is in a strongly localized state (the $IP_{d}$ in the P-doped system
and the $EA_{d}$ in the B-doped system agree remarkably well with
sBNL).

\begin{table*}[t]
\begin{tabular}{c|cccc|cccc}
 & \multicolumn{4}{c|}{P (center)} & \multicolumn{4}{c}{B (center)}\tabularnewline
method & $IP_{d}$ & $EA_{u}$ & $E_{\text{act}}$ & diff. & $IP_{u}$ & $EA_{d}$ & $E_{\text{act}}$ & diff.\tabularnewline
\hline 
sBNL & 4.292 & 1.296 & \textbf{2.996} & 0.000 & 7.711 & 4.323 & \textbf{3.387} & 0.000\tabularnewline
sGW & 4.498 & 1.785 & \textbf{2.713} & -0.283 & 8.565 & 5.540 & \textbf{3.025} & -0.362\tabularnewline
LDA & 3.173 & 2.739 & \textbf{0.434} & -2.562 & 6.123 & 5.591 & \textbf{0.531} & -2.856\tabularnewline
$\Delta$SCF(BNL) & 4.333 & 1.343 & \textbf{2.990} & -0.006 & 7.795 & 4.501 & \textbf{3.294} & -0.093\tabularnewline
$\Delta$SCF(LDA) & 4.449 & 1.546 & \textbf{2.903} & -0.093 & 7.348 & 4.247 & \textbf{3.101} & -0.286\tabularnewline
\end{tabular}

\caption{\label{tbl2:stoGWcompare} Method comparisons for dopant activation
energies (eV) in $D=1$~nm crystal. Differences are taken with respect
to sBNL.}
\end{table*}

\begin{figure}[b]
\includegraphics[width=0.25\linewidth]{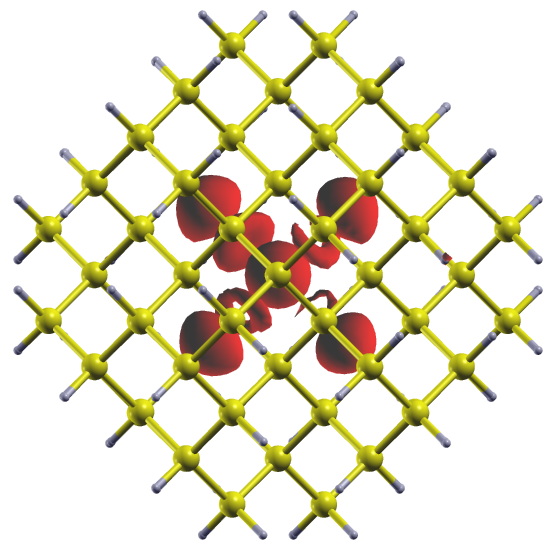}\includegraphics[width=0.25\linewidth]{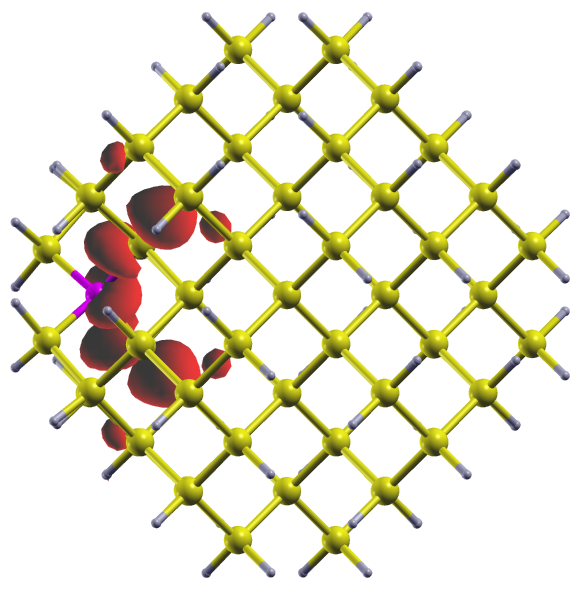}\includegraphics[width=0.25\linewidth]{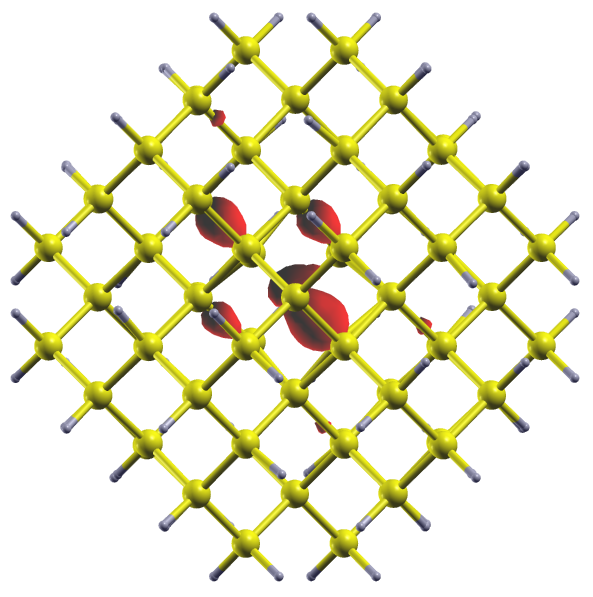}\includegraphics[width=0.25\linewidth]{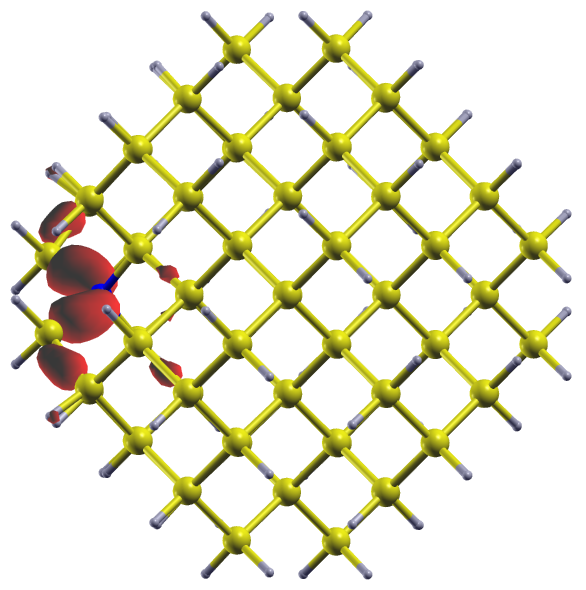}
\caption{\label{fig:4} Charge densities for dopant level in $D=1.4$~nm nanocrystal
for (left to right) P center, P surface, B center, and B surface configurations.
The isosurface is plotted at $20-25\%$ of the maximum value.}
\end{figure}

In the size regime tested ($D=1-2.5$~nm), the defect states from
the P- and B- dopants are strongly localized, which can be observed
in their charge density plots (Fig.~\ref{fig:4}). As a result, the
defect states lie deep in the gap compared to bulk, reflecting their
large ($\ge1$~eV) activation energies. This also leads to an IP
that is roughly independent of NC size for the P-doped system and
to an EA that is roughly independent of NC size for the B-doped system,
consistent with previous theoretical and experimental studies.\citep{Melnikov2004,Chan2008,Cantele2005,Iori2007,Fujii2016}

For the P dopant, when P is in the center of the crystal, the IP hardly
changes with nanocrystal size (the red line that tracks the $IP_{d}$
trend is nearly flat). This can be attributed to the strong electron-impurity
interaction in the confined system that gives rise to the localized
defect state.\citep{Melnikov2004} The size dependence of the activation
energy is therefore almost entirely due to the confinement of the
LUMO in the undoped system. We note a tendency for the activation
energy to decrease slightly when the dopant moves to the surface.
A possible explanation for this is when the dopant is placed near
the surface, its wave function becomes more distorted and less symmetrical,
reducing its Coulomb binding energy and therefore its $IP_{d}$. Calculated
and experimentally measured values of the hyperfine splitting parameter
in P-doped structures support this interpretation.\citep{Chan2008,Fujii2002}

The B dopants can be interpreted with a similar analysis. Because
the dopant level is also localized and does not vary much with crystal
size, the trend in the activation energy is mostly governed by the
confinement of the $IP_{u}$ in the undoped system. However, structural
relaxation plays a greater role due to the smaller size of the B atom
(see Ref.~\onlinecite{Ossicini2006} and Supplementary Material).
Activation energies for the B dopants tend to be higher when the dopant
is near the surface. This likely comes about due to structural relaxation
effects and spin splitting. Spin splitting occurs because the bond
lengths to the neighboring Si atoms are not evenly distributed around
the dopant atom. When B is in the center, its bonds to the neighboring
silicons are almost equivalent. When B is near the surface, the bonds
to the outer Si atoms contract more than the bonds to the inner Si
atoms. This uneven distribution contributes to increased spin splitting
which raises the energy level for the dopant. For example, in the
$D=1$~nm NC, the spin splitting value is $2.842$~eV when B is
at the center and $3.026$~eV when B is near the surface. This interpretation
follows that in another DFT study on B-doped Si NCs.\citep{Ni2017}

\section{Conclusions}

We applied the stochastic BNL approach to study dopant activation
energies for P- and B- doped Si nanocrystals with up to $1600$ electrons.
The stochastic approach reduces the scaling of the exact exchange
from quadratic to linear, enabling its application to experimentally
relevant system sizes. We find excellent agreement with $\Delta\text{SCF}$ and good agreement with stochastic GW for dopant activation
energies using a single range-parameter ($\gamma$) for the stochastic
BNL functional. The difference for stochastic GW could be influenced by the underlying LDA starting point, which erroneously predicts shallow dopant levels. One of the key findings is that shallow dopants in
the bulk become deep dopants under confinement. This has been observed
in previous studies using $\text{\ensuremath{\Delta}SCF}$ at the
LDA level, but we can finally use stochastic BNL to validate these
results at the hybrid functional level.

This study is further significant in that it provides a way to
calculate a self-consistent solution with a quasiparticle model Hamiltonian
at a low computational cost. This quasiparticle Hamiltonian, which is not obtainable through traditional GW methods, can be
further processed and used with methods like time-dependent
DFT and stochastic BSE\citep{Rabani2015} to describe optical excitations for system sizes and complexities beyond current limitations.

\begin{acknowledgments}
The authors thank Helen Eisenberg for valuable discussions and coding
assistance. We acknowledge support from the Center for Computational
Study of Excited State Phenomena in Energy Materials (C2SEPEM) at
the Lawrence Berkeley National Laboratory, which is funded by the
U.S. Department of Energy, Office of Science, Basic Energy Sciences,
Materials Sciences and Engineering Division under Contract No. DE-AC02-05CH11231
as part of the Computational Materials Sciences Program. Computational
resources were provided by the National Energy Research Scientific
Computing Center (NERSC), a U.S. Department of Energy Office of Science
User Facility operated under Contract No. DE-AC02-05CH11231. We would
also like to thank the computational resources provided by XSEDE under
Project No. TG-CHE170058. RB gratefully thanks the Israel-US
Binational Science Foundation (BSF) Grant No. 2018368.
\end{acknowledgments}

\bibliographystyle{apsrev4-1}

%

\end{document}